\documentstyle[]{article}
\newlength{\aivwidth}   
\newlength{\tmpwidth}   

\title{ Einstein-Infeld-Hoffman method and soliton dynamics in 
        a parity noninvariant system }
\author{Jacek Dziarmaga  \\
        Department of Mathematical Sciences,\\
        University of Durham, South Road, Durham, DH1 3LE,\\ 
        United Kingdom.\\
        E-mail address: J.P.Dziarmaga@durham.ac.uk 
        \thanks{On leave of absence from Institute of Physics,
        Jagellonian University, Krak\'ow, Poland} }
\date{February 20, 1996}
\begin{document}
\maketitle
   \begin{abstract}
  We consider slow motion of a pointlike topological defect (vortex) in 
the nonlinear Schrodinger equation minimally coupled to Chern-Simons gauge 
field and subject to external uniform magnetic field. It turns out that a 
formal expansion of fields in powers of defect velocity yields only the 
trivial static solution. To obtain a nontrivial solution one has to treat 
velocities and accelerations as being of the same order. We assume that
acceleration is a linear form of velocity. The field equations linearized 
in velocity uniquely determine the linear relation. It turns out that the 
only nontrivial solution is the cyclotron motion of the vortex together 
with the whole condensate. This solution is a perturbative approximation to
the center of mass motion known from the theory of magnetic translations.
   \end{abstract}
\vspace*{1cm}
DTP-95/71\\
\vspace*{1cm}

\section{Introduction}

  In this paper we are going to discuss the reliability of the 
Einstein-Infeld-Hoffman (EIH) method \cite{aih}, which was developed in the 
theory of general relativity for analysis of the motion of point 
sources of gravitational field, to dynamics of point-like topological defects 
in parity noninvariant systems. The essence of the EIH method, which
remains relevant in the present context, can be outlined as follows.
Let us assume that there is a characteristic velocity $v$ in the model.
In a relativistic model $v$ is the velocity of light while in 
a nonrelativistic model $v$ can be say a sound velocity. To consider
time evolution which is slow as compared to the characteristic velocity, 
one can develope a perturbative expansion with $1/v$ as an expansion
parameter. The fields are expanded in powers of $1/v$ and the 
time $t$ is replaced by a rescaled time $\tau$, $t=v\tau$.
The time derivatives are rescaled as 
$\partial_{t}=\frac{1}{v}\partial_{\tau}$.
The $n$-th order time derivative of a given quantity is formally
$n$ orders of magnitude smaller than the quantity itself. In particular
acceleration is always negligible as compared to velocity.
However in a parity noninvariant system like a system in an external
uniform magnetic field one can imagine a soliton performing a
cyclotron kind of motion. In the cyclotron motion acceleration
is always a linear form of velocity so that this kind of
motion can not be described within the formal expansion in powers
of $1/v$.

  The standard approach to the dynamics of point particles interacting
with fields is to take in the first step arbitrary particle trajectories
and find from field equations the fields produced by the assumed
currents. In this first step the parameter $1/v$ can
be employed to calculate the fields an expansions in powers of
particle velocities. The second step is to substitute the calculated
fields into particles equations of motion. In this way
one obtains purely mechanical equations which can be solved
to obtain particle trajectories. Such a method was applied to derive
the Darwin Lagrangian in electrodynamics \cite{jackson} or in the context
of general relativity \cite{aih} to mention only the most important
cases. 

  It turns out that in the case of topological defects
such an off-shell calculation is not possible. One can try to
find solutions of field equations for a given soliton 
trajectory but it turns out that a regular solution exists
but only for a very special trajectory which is thus the 
solution of the soliton dynamics problem. This property has 
been observed first in the case of relativistic Chern-Simons
vortices \cite{jd} and independently for relativistic 
membranes \cite{arodz}. The origin of the problem in the case of 
solitons can be traced back to the fact that, unlike for point particles 
interacting with fields, there is no a priori 
equation of motion for the topological defects. Unlike point
particles solitons are composed of the same fields as the 
fields which mediate interactions between them. Their dynamics is
implicit in the field equations.
Our calculation
in Section 3 provides another example. Its advantage is that
there is an exact solution at hand to be compared with the perturbative
result.

\section{Model and boost in an external magnetic field}

  Let us consider the model \cite{ezawa}, which is a field-theoretical 
description of the quantum Hall effect of polarized electrons. The electrons 
are described in terms of bosonic fields fermionized by coupling to 
the auxillary Chern-Simons fields $a_{\mu}$, 
\begin{eqnarray}\label{g20}
L=\frac{\kappa}{2}\varepsilon^{\gamma\alpha\beta}
                        a_{\gamma}\partial_{\alpha}a_{\beta}  
 +\frac{i}{2}\{\Psi^{\star}(D_{0}\Psi)-\Psi(D_{0}\Psi)^{\star}\}
                    -\frac{1}{2m}(D_{k}\Psi)^{\star}(D_{k}\Psi)
                    -\frac{\lambda}{2}(\rho_{0}-\Psi^{\star}\Psi)^{2} \;\;.
\end{eqnarray}
$\rho_{0}$, which is a condensate density, is related to the external
magnetic field by $\rho_{0}=-\kappa B/e$.  In what follows we assume
$\kappa$ to be negative. The covariant derivative couples the scalar 
field to both the external electromagnetic field $A_{\mu}$ and to 
the Chern-Simons field $a_{\mu}$, $D_{\mu}=\partial_{\mu}-iea_{\mu}-ieA_{\mu}$. 
The Greek indices run over space-time indices $0,1,2$ while the Latin
indices denote planar coordinates $1,2$. Our convention is 
$\varepsilon^{012}=+1$ and we assume the signature $(+,-,-)$. In the 
formulation (\ref{g20}) 
the ground state of the theory in the external magnetic field B,
\begin{eqnarray}\label{g25}
&& A_{0}(t,\vec{x})=0 \;\;,\nonumber\\
&& A_{k}(t,\vec{x})=\frac{1}{2}B\varepsilon_{kl}x^{l} \;\;,
\end{eqnarray}
is the uniform condesate $\Psi^{\star}\Psi=\rho_{0}$ with the external
magnetic field being screened by the Chern-Simons field,
$a_{k}=-A_{k},\; a_{0}=0$. Such a ground state admits topological vortex 
excitations \cite{ezawa,bh1,bh2}.

  The theory (\ref{g20}) is a Galilean invariant system in an
external magnetic field. As such it has the following symmetry.
If the set of fields $\Psi(t,\vec{x}),a_{\mu}(t,\vec{x})$ is a solution
of the model (\ref{g20}), then the following boosted fields are
also solutions of the model
\begin{eqnarray}\label{g70}
&&\tilde{\Psi}(t,\vec{x})=\Psi[t,\vec{x}-\vec{R}(t)] e^{i\chi_{B}}
                                                             \;\;,\nonumber\\ 
&&\tilde{a}_{0}(t,\vec{x})=a_{0}[t,\vec{x}-\vec{R}(t)]
                -\dot{R}^{k}a_{k}[t,\vec{x}-\vec{R}(t)]  \;\;,\nonumber\\
&&\tilde{a}_{k}(t,\vec{x})=a_{k}[t,\vec{x}-\vec{R}(t)]  \;\;,\nonumber\\
&&\chi_{B}=-\frac{1}{2}m(\dot{R}^{k}\dot{R}^{k})t+m\dot{R}^{k}x^{k}
  +e\int_{t_{0}}^{t}d\tau\;\dot{R}^{k}(\tau)A_{k}[\vec{x}-\vec{R}(\tau)]\;\;.
\end{eqnarray}
provided that the trajectory $\vec{R}(t)$ satisfies the equation
of motion
\begin{equation}\label{g75}
m\ddot{R}^{k}=-eB\varepsilon^{kl}\dot{R}^{l} \;\;.
\end{equation}
The last equation is simply the equation of motion of a planar electron
in uniform magnetic field. Its solution is the cyclotron motion with
the cyclotron frequency $\omega_{c}=eB/m$. If an unboosted solution
is say the uniform condesate, then the solution after the boost
is a condesate performing a cyclotron motion. If the unboosted
solution contained a vortex then after the boost the vortex would
move together with the condensate.

\section{ Slow vortex motion in perturbative approximation }

  In this section we are going to describe a perturbative scheme
which can be applied in investigations of slow motions of topological
defects in parity noninvariant systems. As the perturbative
method is quite general, we think it may find applications in some
other models. The method has been already described in our earlier
article \cite{jd} devoted to the dynamics of relativistic self-dual
Chern-Simons vortices. However, the calculation in \cite{jd}
is relatively complicated and what is more there is no exact
solution at hand to be compared with the perturbative result.
The magnetic boost provides us with a nice example of exact solution, which 
confirms the perturbative result and the
perturbative method as such. The perturbative calculation shows
also uniqueness of the magnetic boost. 

   Before we describe the perturbative calculation let us make a small
rearrangement in the model (\ref{g20}). Namely, let us replace
$a_{\mu}+A_{\mu}=\hat{a}_{\mu}$. This replacement and the use of the 
definition $\rho_{0}=-\frac{\kappa B}{e}$ leads \cite{bh2} to the equivalent 
version of the model (\ref{g20})
\begin{eqnarray}\label{s20}
L=\frac{\kappa}{2}\varepsilon^{\gamma\alpha\beta}
  a_{\gamma}\partial_{\alpha}a_{\beta}-e\rho_{0}a_{0}
 +\frac{i}{2}\{\Psi^{\star}(D_{0}\Psi)-\Psi(D_{0}\Psi)^{\star}\}
                    -\frac{1}{2m}(D_{k}\Psi)^{\star}(D_{k}\Psi)
                    -\frac{\lambda}{2}(\rho_{0}-\Psi^{\star}\Psi)^{2} \;\;,
\end{eqnarray}
where we have already neglected the hats. The covariant derivative is 
simplified to $D_{\mu}=\partial_{\mu}-iea_{\mu}$. The advantage of this 
formulation  is that there is just the Chern-Simons gauge field to be 
handled with.  The external magnetic field is replaced by the uniform 
background charge density now.
 
   We concentrate on the model (\ref{s20}) from now on. The field
equations of the model are
\begin{eqnarray}\label{s100}
&& i\partial_{t}\Psi+ea_{0}\Psi+\frac{1}{2m}D_{k}D_{k}\Psi
                   +\lambda(\rho_{0}-\Psi^{\star}\Psi)\Psi=0
\;\;,\nonumber \\
&& \kappa\varepsilon_{kl}\partial_{k}a_{l}+e(\rho-\rho_{0})=0 \;\;, 
\nonumber \\
&& \kappa\varepsilon^{k\alpha\beta}\partial_{\alpha}a_{\beta}+eJ^{k}=0 \;\;,
\end{eqnarray}
where $\rho=\Psi^{\star}\Psi$ is the particle density and the current is 
\begin{equation}\label{s110}
J^{k}=\frac{i}{2m}\{\Psi(D_{k}\Psi)^{\star}-\Psi^{\star}(D_{k}\Psi)\}\equiv
\frac{\rho}{m}(\partial_{k}\chi-ea_{k}) \;\;.
\end{equation}
$\chi$ is the phase of the scalar field, $\Psi=\sqrt{\rho}\exp i\chi\;$.
The model admits the uniform condensate solution $\Psi=\sqrt{\rho_{0}}$,
$a_{\mu}=-A_{\mu}$. Because there is a nonvanishing condensate,
the model also admits topological vortex solutions. The Ansatz
for a vortex solution with the winding number minus one can be taken as
\begin{eqnarray}\label{v.10}
&& \Psi^{(0)}(\vec{x})=[\rho(r)]^{1/2}e^{-i\theta} \;\;,\nonumber\\
&& a^{(0)}_{\theta}(\vec{x})=A(r)    \;\;,\nonumber\\
&& a^{(0)}_{0}(\vec{x})=A_{0}(r)                  \;\;,\nonumber\\
&& a^{(0)}_{r}(\vec{x})=0 \;\;.
\end{eqnarray}
The function $\rho(r)$ interpolates between $\rho(0)=0$ and
$\rho(\infty)=\rho_{0}$. The gauge potential vanishes at the origin,
$A(0)=0$, and tends to a pure gauge at infinity, $A(r)\approx-\frac{1}{er}$.
In the Bogomol'nyi limit, $\lambda=\frac{e^{2}}{\mid\kappa\mid m}$, the 
equations fulfilled by the profile functions in Eq.(\ref{v.10}) can be 
derived \cite{bh1,bh2} from
\begin{eqnarray}\label{v.20}
&& \nabla^{2}\ln\rho=-\frac{2e^{2}}{\kappa}(\rho-\rho_{0}) \;\;,\nonumber\\
&& A(r)=-\frac{1}{er}+\frac{\rho'}{2e\rho}                 \;\;,\nonumber\\
&& A_{0}(r)=\frac{e}{2m\kappa}(\rho-\rho_{0})              \;\;.
\end{eqnarray}  
The primes denote derivatives with respect to $r$, $"'"\equiv\frac{d}{dr}$.
Once the solution of the first equation in (\ref{v.20}) is known 
the functions $A(r)$ and $A_{0}(r)$ can be expressed through 
$\rho(r)$.

  The aim of the perturbative calculation is to find an approximate 
trajectory of a vortex in the limit of slow motion. The perturbative 
method consists of two main ingredients. The first of them is rather 
classic. As we are interested in slowly moving vortices,
we expand all the quantities in powers of vortex velocity $\dot{R}^{k}(t)$.
The zero order approximation to a moving vortex solution would be just
\begin{eqnarray}\label{p.10}
&&\bar{\Psi}(t,\vec{x})\equiv\Psi^{(0)}[\vec{x}-\vec{R}(t)] \;\;,\nonumber\\
&&\bar{a}_{\mu}(t,\vec{x})=a^{(0)}_{\mu}[\vec{x}-\vec{R}(t)] \;\;.
\end{eqnarray}   
It is a solution to field equations but only when we neglect all the terms 
in field equations linear (or higher order) in vortex velocities. 
An exact solution $\Psi(t,\vec{x}),a_{\mu}(t,\vec{x})$, if exists, differs 
from the fields in Eqs.(\ref{p.10}), 
\begin{eqnarray}\label{p.20}
&&\Psi(t,\vec{x})=\bar{\Psi}(t,\vec{x})+\psi(t,\vec{x}) \;\;,\nonumber\\
&&a_{\mu}(t,\vec{x})=\bar{a}_{\mu}(t,\vec{x})+u_{\mu}(t,\vec{x}) \;\;.
\end{eqnarray}
The deviations $\psi,u_{\mu}$ are at least of first order in
vortex velocity. The field equations (\ref{s100}), when linearized
both in the vortex velocity and in the deviations $\psi,u_{\mu}$,
become a set of inhomogenous differential equations
\begin{eqnarray}\label{p.30}
&&i\partial_{t}\psi+e\bar{a}_{0}\psi+e\bar{\Psi}u_{0}+
   \frac{1}{2m}[\bar{D}_{k}\bar{D}_{k}\psi
   -2ie(\bar{D}_{k}\bar{\Psi})u_{k}-ie\bar{\Psi}(\bar{D}_{k}u_{k})]
+\lambda\rho_{0}\psi-2\lambda\bar{\Psi}^{\star}\bar{\Psi}\psi
-\lambda\bar{\Psi}^{2}\psi^{\star}
                                   =-i\partial_{t}\bar{\Psi} \;\;,\nonumber\\
&&\kappa\varepsilon_{kl}\partial_{k}u_{l}
                     +e(\bar{\Psi}^{\star}\psi+\bar{\Psi}\psi^{\star})=0 
\;\;,\nonumber\\
&&\kappa\varepsilon^{k\alpha\beta}\partial_{\alpha}u_{\beta}
  -e^{2}\bar{\Psi}^{\star}\bar{\Psi}u_{k}
  +e(\partial_{k}\bar{\chi}-e\bar{a}_{k})
    (\bar{\Psi}^{\star}\psi+\bar{\Psi}\psi^{\star})  
   =\kappa\varepsilon^{kl}\partial_{t}\bar{a}_{l} \;\;,
\end{eqnarray}
where $\bar{D}_{k}=\partial_{k}-ie\bar{a}_{k}$.
The deviations $\psi,u_{\mu}$ can be expanded 
in powers of vortex velocity components $\dot{R}^{k}(t)$.
The leading terms in the expansion read
\begin{eqnarray}\label{p.40}
&&\psi(t,\vec{x})=\dot{R}^{k}(t)\psi^{(k)}[\vec{x}-\vec{R}(t)]
                                                    \;\;,\nonumber\\
&&u_{\mu}(t,\vec{x})=\dot{R}^{k}(t)u_{\mu}^{(k)}[\vec{x}-\vec{R}(t)] \;\;.
\end{eqnarray}

  Before we substitute the Ansatz (\ref{p.40}) to the linearized
equations (\ref{p.30}), we have to introduce the second and key ingredient of 
the perturbative method. Namely we allow the vortex acceleration to be 
manifestly linear in velocity   
\begin{equation}\label{p.50}
\ddot{R}^{k}(t)=\omega^{kl}\dot{R}^{l}(t)+O(\dot{R}^{k}\dot{R}^{k}) \;\;.
\end{equation}
The matrix $\omega$ is time-independent and velocity-independent.
In other words we are going to consider accelerations as being
of the same order in magnitude as velocities.
 
  The assumption that the acceleration and the velocity are of the same 
order contradicts customary claims in the literature. The usual
reasoning, which can be traced back to the Einstein-Infeld-Hoffman method
in general relativity \cite{aih}, is as follows. We assume there is   
a characteristic velocity in the model, say $v$. To consider
slow time-evolution, we develope a perturbative expansion
in the parameter $1/v$, which we consider to be small. The perturbative
expansion has two ingredients. First of all the fields are expanded
around boosted static solutions in the powers of $1/v$. Second, the real 
time $t$ is replaced by a rescaled time $\tau$, $t=v\tau$. The time
derivatives are then rescaled as $\partial_{t}=\frac{1}{v}\partial_{\tau}$.
For a given quantity $Q$, its $n$-th time derivative is formally $n$   
orders smaller than the quantity itself, 
$\partial_{t}^{n}Q=v^{-n}\partial_{\tau}^{n}Q$.
In particular in this perturbative scheme the acceleration is
always regarded to be negligible as compared to velocity or in other
words the matrix $\omega^{kl}$ in Eq.(\ref{p.50}) is implicitly assumed
to be zero. Thus the commonly accepted perturbative scheme rules
out any solution like a motion along circular orbit. 
     
We still do not have a satisfactory perturbative scheme to describe
eventual vortex motion with radiation in which the back-reaction could be
self-consistently taken into account. Thus in this paper we will 
restrict to nonradiative trajectories. When there is no radiation
the energy of the vortex must be conserved. For a single vortex in an 
uniform condensate, translational invariance implies that its acceleration 
must be perpendicular to velocity 
\begin{equation}\label{p.60}
\omega^{kl}=\omega\varepsilon^{kl} \;\;,
\end{equation}
where $\omega$ is a constant. We do not know the value of the 
constant. It has to be fixed by a solvability condition.
 
  Let us then try to solve the Eqs.(\ref{p.30}). The examination
of sources on the RHS's of Eqs.(\ref{p.30}) 
\begin{eqnarray}\label{p.65}
&&-i\partial_{t}\bar{\Psi}=
  \frac{i\rho'}{2\rho}[\dot{R}^{1}\cos\theta+\dot{R}^{2}\sin\theta]
  +\frac{\rho^{1/2}}{r}[-\dot{R}^{1}\sin\theta+\dot{R}^{2}\cos\theta]
                                                         \;\;,\nonumber\\
&&\partial_{t}\bar{a}_{1}=
  [\dot{R}^{1}\cos\theta+\dot{R}^{2}\sin\theta]A'\sin\theta
  +[-\dot{R}^{1}\sin\theta+\dot{R}^{2}\cos\theta]\frac{A}{r}\cos\theta
                                                         \;\;,\nonumber\\
&&\partial_{t}\bar{a}_{2}=
-[\dot{R}^{1}\cos\theta+\dot{R}^{2}\sin\theta]A'\cos\theta
-[-\dot{R}^{1}\sin\theta+\dot{R}^{2}\cos\theta]\frac{A}{r}\sin\theta 
\end{eqnarray}
shows that without loss of generality we can adopt the following Ansatz 
for the first order field deviations (we assume the gauge in which the 
deviation of the scalar field's phase is zero)
\begin{eqnarray}\label{p.70}
&&\dot{R}^{k}f^{(k)}(\vec{x})=
  \rho^{1/2}(r)s(r)[-\dot{R}^{1}\sin\theta+\dot{R}^{2}\cos\theta] 
                                                          \;\;,\nonumber\\ 
&&\dot{R}^{k}u^{(k)}_{0}(\vec{x})=
      a(r)[-\dot{R}^{1}\sin\theta+\dot{R}^{2}\cos\theta]  \;\;,\nonumber\\
&&\dot{R}^{k}u^{(k)}_{r}(\vec{x})=
       b(r)[\dot{R}^{1}\cos\theta+\dot{R}^{2}\sin\theta]  \;\;,\nonumber\\ 
&&\dot{R}^{k}u^{(k)}_{\theta}(\vec{x})=
       c(r)[-\dot{R}^{1}\sin\theta+\dot{R}^{2}\cos\theta] \;\;.
\end{eqnarray}
The substitution of the above Ansatz to Eqs.(\ref{p.30}) yields
\begin{eqnarray}\label{p.80}
&&s''+\frac{s'}{r}-\frac{s}{r^{2}}+\frac{\rho'}{\rho}(s'-c)+2a
                                    -4\rho s=\frac{2}{r} \;\;,\nonumber\\
&&b'+\frac{b}{r}-\frac{c}{r}+\frac{\rho'}{\rho}(b-\frac{s}{r})+2\omega s=
                                        -\frac{\rho'}{\rho} \;\;,\nonumber\\
&&c'+\frac{c}{r}-\frac{b}{r}-2\rho s=0                      \;\;,\nonumber\\
&&a'-\omega b-\rho' s-\rho c=\frac{A}{r}                    \;\;,\nonumber\\
&&\frac{a}{r}-\omega c-\rho b=A' \;\;.
\end{eqnarray}
We use from now on the rescaled units in which all the constants
$m,(-\kappa),\rho_{0}$ and $e$ are set equal to 1. We have also restricted 
to the
Bogomol'nyi limit, which in the rescaled units corresponds to $\lambda=1$.
It is by no means necessary but it makes the formulas more compact.
In particular the second equation in the set (\ref{v.20}), which holds
only in the Bogomol'nyi limit, can serve for many simplifications.
 
The second equation in the set
(\ref{p.80}) is not independent, it can be derived from the last three
equations. The last equation in the set (\ref{p.80})
can be used to express $a(r)$ by other functions
\begin{equation}\label{p.90}
a=\omega rc+r\rho b+A'r \;\;.
\end{equation}
Once $a(r)$ is expressed like in Eq.(\ref{p.90}), it can be eliminated
from the first, third and fourth equation in the set (\ref{p.80}).
Finally we are left with only three independent equations
\begin{eqnarray}\label{p.100}
&&b'+\frac{b}{r}-\frac{c}{r}+2\omega s
                     +\frac{\rho'}{\rho}[b-\frac{s}{r}+1]=0 \;\;,\nonumber\\
&&c'+\frac{c}{r}-\frac{b}{r}-2\rho s=0                      \;\;,\nonumber\\
&&s''+\frac{s'}{r}-\frac{s}{r^{2}}+\frac{\rho'}{\rho}(s'-c)-4\rho s
              +2r\rho b+2\omega rc=2r(\frac{1}{r^{2}}-A') \;\;.
\end{eqnarray}
The regular asymptotes close to $r=0$ are
\begin{eqnarray}\label{p.110}
&&s(r)\approx (1+\alpha) r+... \;\;,\nonumber\\
&&b(r)\approx \alpha+... \;\;,\nonumber\\
&&c(r)\approx \alpha+... \;\;,
\end{eqnarray}
where the higher order terms in $r$ were neglected. There is only one
free parameter $\alpha$ in the asymptote.
 
  To find out the asymptote at infinity we go back to Eqs.(\ref{p.80}).
At infinity $\rho\approx 1$, up to terms decaying exponentially.
The asymptotic form of Eqs.(\ref{p.80}) is
\begin{eqnarray}\label{p.120}
&&s''+\frac{s'}{r}-\frac{s}{r^{2}}+2a-4s=\frac{2}{r} \;\;,\nonumber\\
&&b'+\frac{b}{r}-\frac{c}{r}+2\omega s= 0             \;\;,\nonumber\\
&&c'+\frac{c}{r}-\frac{b}{r}-2s= 0                    \;\;,\nonumber\\
&&a'+\frac{1}{r^{2}}=\omega b+c                       \;\;,\nonumber\\
&&\frac{a}{r}-\frac{1}{r^{2}}=b+\omega c             \;\;.
\end{eqnarray}
This set of asymptotic equations implies the following
\begin{eqnarray}\label{p.130}
&&s''+\frac{s'}{r}-\frac{s}{r^{2}}-4s+2a=\frac{2}{r} \;\;,\nonumber\\
&&a''+\frac{a'}{r}-\frac{a}{r^{2}}-2(1-\omega^{2})s=0 \;\;.\nonumber\\
\end{eqnarray}
Once $s(r)$ and $a(r)$ are known, one can find $b(r)$ and $c(r)$
solving with respect to them the last two equations in the set (\ref{p.120}). 
The solution is unique provided that $\omega^{2}\neq1$. 

One can diagonalize Eqs.(\ref{p.130}). The eigenvalues turn out to be
\begin{equation}\label{p.140}
\lambda_{1,2}=-2(1\stackrel{+}{-}\omega) \;\;.
\end{equation}
If $\omega^{2}>1$, one of the eigenvalues is positive and the other
is negative. The asymptote must contain a linear combination of two Bessel
functions and of two modified Bessel functions. Thus there is one divergent
mode and two long-range unnormalisable modes to be removed. But we have
only two adjustable constants, $\alpha$ in the asymptote at the origin
(\ref{p.110}) and $\omega$. Thus there is no normalisable solution for
$\omega^{2}>1$.
              
If $\omega^{2}<1$, the general asymptote is a combination of two
exponentially divergent modes and two normalisable modes. Thus
the two constants $\omega,\alpha$ might happen to be sufficient to remove
the two divergent modes. Numerical analysis shows that it does not
happen for $-1<\omega<1$, however.
 
 Thus we must restrict our attention to the case $\omega^{2}=1$. 
For $\omega=+1$ the asymptotic equations (\ref{p.120}) are solved by
\begin{eqnarray}\label{p.150}
&&a(r)\approx\frac{1}{r}+2\varepsilon r                 \;\;,\nonumber\\ 
&&s(r)\approx\beta_{1}\frac{e^{2r}}{\sqrt{2r}}
     +\beta_{2}\frac{e^{-2r}}{\sqrt{2r}}+\varepsilon r  \;\;,\nonumber\\ 
&&b(r)\approx-c(r)\approx \gamma r^{-2}+\frac{\varepsilon}{2}r^{2} \;\;.  
\end{eqnarray}
The two constants $\beta_{1}$ and $\varepsilon$ must be tuned to zero
with just one parameter $\alpha$ in the asymptote close to the origin.
There is no regular solution for $\omega=+1$.

For $\omega=-1$ the asymptote is 
\begin{eqnarray}\label{p.155}
&&a(r)\approx\frac{1}{r}+\frac{2\sigma}{r}                 \;\;,\nonumber\\
&&s(r)\approx\beta_{1}\frac{e^{2r}}{\sqrt{2r}}
     +\beta_{2}\frac{e^{-2r}}{\sqrt{2r}}+\frac{\sigma}{r}  \;\;,\nonumber\\
&&b(r)=-c(r)\approx \delta                                 \;\;.
\end{eqnarray}
For nonzero $\delta$ the gauge fields' deviations become
$\dot{R}^{k}a^{(k)}_{l}=\delta\dot{R}^{l}$, compare with 
Eq.(\ref{p.70}).  
Once the constant $\beta_{1}$ is tuned to zero with the help of the 
free parameter $\alpha$, the asymptote of the electric current becomes
$J^{k}=\delta e^{2}\rho_{0}\dot{R}^{k}$. Thus at any given
moment of time the vortex velocity is parallel to the condensate current.
What is more, for $\omega=-1$ one can give an explicit unique solution to
Eqs.(\ref{p.100}), namely $b(r)=c(r)=-1$ and $s(r)=0$. According to
Eq.(\ref{p.90})
\begin{equation}\label{p.175}
a(r)=r[1-\rho(r)+A'(r)]=-A(r) \;\;.
\end{equation}
The last equality holds thanks to the Gauss law. Thus the constant 
$\delta$ is fixed as $\delta=-1$. The vortex moves together with
the negatively charged condensate. 

   Note that $\omega=-1$ so the condesate moves along a cyclotron orbit, 
$\ddot{R}^{k}=-\varepsilon^{kl}\dot{R}^{l}$. The solution is a perturbative 
approximation to the exact deformed Galilean boost. Such a solution could not 
be obtained in a perturbative calculation if accelerations were neglected as 
compared to velocities. At the same time, the analysis shows that
the deformed Galilean boost is the only solution which can be obtained
by a perturbative expansion in velocities. In principle this does
not exclude possibility of a dissipative noncyclotron motion.

\paragraph{Aknowledgement.}
I would like to thank Professor Andrzej Staruszkiewicz for a discussion
which initiated this work. This research was supported in part by the KBN 
grant 2 P03B 085 08 and by Foundation for Polish Science and UKPPARC.

\end{document}